\documentclass[prl,reprint,superscriptaddress]{revtex4-1}
\usepackage{graphicx}
\usepackage{dcolumn}
\usepackage{bm}
\usepackage[english]{babel}
\usepackage{hyperref}
\addto\captionsenglish{}
\newcommand\myhash{\scalebox{0.8}{\raisebox{0.4ex}{\#}}}
\hypersetup{colorlinks=true, urlcolor=blue, citecolor=blue, linkcolor=blue, anchorcolor=blue}  

\begin{document}


\title{Loss of Polarization in Collapsing Beams}



\author{Gauri Patwardhan}
\affiliation{Department of Applied Physics and Applied Mathematics, Columbia University, New York, NY 10027, USA}
\affiliation{School of Applied and Engineering Physics, Cornell University, Ithaca, NY 14853, USA}

\author{Xiaohui Gao}%
\affiliation{Department of Applied Physics and Applied Mathematics, Columbia University, New York, NY 10027, USA}
\author{Amir Sagiv}
\affiliation{Department of Applied Mathematics, Tel Aviv University, Tel Aviv 6997801, Israel}%

\author{Avik Dutt}%
\affiliation{Department of Electrical Engineering, Columbia University, New York, NY 10027, USA}
\affiliation{School of Electrical and Computer Engineering, Cornell University, Ithaca, NY 14853, USA}

\author{Jared Ginsberg}
\affiliation{Department of Applied Physics and Applied Mathematics, Columbia University, New York, NY 10027, USA}%
\author{Adi Ditkowski}
\affiliation{Department of Applied Mathematics, Tel Aviv University, Tel Aviv 6997801, Israel}%
\author{Gadi Fibich}
\affiliation{Department of Applied Mathematics, Tel Aviv University, Tel Aviv 6997801, Israel}%
\author{Alexander L. Gaeta}
\email{a.gaeta@columbia.edu}
\affiliation{Department of Applied Physics and Applied Mathematics, Columbia University, New York, NY 10027, USA}


\date{\today}

\begin{abstract}
We show theoretically and demonstrate experimentally that collapsing elliptically-polarized laser beams experience a nonlinear ellipse rotation that is highly sensitive to small fluctuations in the input power. For arbitrarily small fluctuations in the input power and after a sufficiently large propagation distance, the polarization angle becomes uniformly distributed in [0, 2$\pi$] from shot-to-shot. We term this novel phenomenon “loss of polarization.” We perform experiments in fused-silica glass, nitrogen gas and water, and observe a significant increase in the fluctuations of the output polarization angle for elliptically-polarized femtosecond pulses as the power is increased beyond the critical power for self-focusing. We also show numerically and confirm experimentally that this effect is more prominent in the anomalous group-velocity dispersion (GVD) regime compared to the normal-GVD regime due to the extended lengths of the filaments for the former. Such effects could play an important role in intense-field light-matter interactions in which elliptically-polarized pulses are utilized.
\end{abstract}

\pacs{42.65.-k, 42.65.Jx, 42.25.Ja, 42.25.-p}

\maketitle
Optical beam collapse occurs when a laser beam with a power greater than a certain critical power $P_{cr}$ propagates through a transparent medium and undergoes self-focusing \cite{fibichSelfFocusingPerturbedUnperturbed1999,fibichCriticalPowerSelffocusing2000,gaetaCatastrophicCollapseUltrashort2000,boydSelffocusingPresentFundamentals2008,fibichNonlinearSchodingerEquation2015}. At higher powers, competing effects such as plasma defocusing arrest the collapse, leading to the formation of laser filaments \cite{couaironFemtosecondFilamentationTransparent2007,kandidovFilamentationHighpowerFemtosecond2009,shimFilamentationAirUltrashort2011} that can confine light over distances much longer than the diffraction length \cite{rodriguezKilometerrangeNonlinearPropagation2004}. Self-focusing and laser filamentation are important for applications in atmospheric remote sensing \cite{kasparianWhiteLightFilamentsAtmospheric2003,kasparianPhysicsApplicationsAtmospheric2008}, light detection and ranging (LIDAR) \cite{rodriguezKilometerrangeNonlinearPropagation2004,hauriGenerationIntenseCarrierenvelope2004}, high-harmonic generation (HHG) \cite{negroFilamentationbasedPolarizationPulse2011,negroPolarizationPulseShaping2010,popmintchevBrightCoherentUltrahigh2012}, pulse compression \cite{hauriGenerationIntenseCarrierenvelope2004}, and terahertz generation \cite{ohIntenseTerahertzGeneration2013}. Additionally, collapsing waves are of universal interest because of their relevance not only in optics but also in a wide variety of fields, e.g., in Bose-Einstein condensation, surface waves dynamics, plasma physics, and Ginzburg-Landau equations \cite{robinsonNonlinearWaveCollapse1997,kivsharSelffocusingTransverseInstabilities2000,pitaevskiiDynamicsCollapseConfined1996,turitsynNonstableSolitonsSharp1993}. 

Through the process of self-phase modulation, the acquired nonlinear phase shift of collapsing beams becomes large and highly sensitive to small fluctuations in the input power, as predicted theoretically \cite{merleUniquenessContinuationProperties1992,fibichContinuationsNonlinearSchrodinger2011} and demonstrated experimentally \cite{shimLossPhaseCollapsing2012}. Furthermore, as the collapsing beam evolves into a filament, the sensitivity of the nonlinear phase shift to small fluctuations increases with propagation disce, so that ultimately, the nonlinear phase shift becomes uniformly distributed in [0,2$\pi$] \cite{sagivLossPhaseUniversality2017}. As a result of this “loss of phase”, the interference between post-collapse beams becomes chaotic \cite{fibichContinuationsNonlinearSchrodinger2011,shimLossPhaseCollapsing2012,sagivLossPhaseUniversality2017,mlejnekOpticallyTurbulentFemtosecond1999,bergeMultipleFilamentationTerawatt2004}.

While the effects of beam collapse on the electric-field amplitude and phase have been extensively investigated \cite{fibichContinuationsNonlinearSchrodinger2011,shimLossPhaseCollapsing2012,varmaTrappingDestructionLongRange2008a,fibichControlMultipleFilamentation2004,haoOptimizationMultipleFilamentation2007,pointSuperfilamentationAir2014}, limited work exists on the study of the polarization of beams undergoing wave collapse. Most of the work studies the effects of polarization on beam collapse \cite{shiInfluenceLaserPolarization2016,panovFilamentationArbitraryPolarized2013,fibichDeterministicVectorialEffects2001,fibichVectorialRandomEffects2001,fibichMultipleFilamentationCircularly2002,fibichSelffocusingCircularlyPolarized2003}. However, the change in the beam’s polarization itself as a result of its collapse remains largely unexplored. Since several applications of laser filamentation including HHG, THz generation and supercontinuum generation are polarization sensitive \cite{youMechanismEllipticallyPolarized2013,dharmadhikariGenerationPolarizationdependentSupercontinuum2015,rostamiDramaticEnhancementSupercontinuum2016}, investigating the polarization state of collapsing beams is crucial \cite{kolesikPolarizationDynamicsFemtosecond2001,phuxuanPolarizationPicosecondLight1983,petitPolarizationDependencePropagation2000}. In some of the studies, molecular alignment and delayed birefringence acting on the probe were investigated \cite{yuanPulsePolarizationEvolution2015,yuanMeasurementBirefringenceFilament2011,kosarevaPolarizationRotationDue2010}. In case of self-induced polarization rotation of the pump, direct measurements \cite{rostamiPolarizationEvolutionUltrashort2015a} using a rotating polarizing cube and indirect measurements \cite{sheinfuxMeasuringStabilityPolarization2012} using femtosecond laser-induced periodic surface structures (FLIPSS) have observed moderate rotations of the polarization angle pre- and post- collapse.  The fluctuations in polarization rotation in these studies, however, were obscured by averaging over multiple shots or pulse periods, and the increase of the fluctuations with propagation distance at powers significantly above $P_{cr}$ was not revealed. Additionally, theoretical investigations based on these observations have not been performed.

In this Letter, we theoretically predict and experimentally demonstrate an effect which we term “loss of polarization.” We show that, when an elliptically-polarized input beam undergoes filamentation, its nonlinear ellipse rotation can become highly sensitive to fluctuations in the input power. Hence, its output polarization becomes random. We show the universality of the loss of polarization effect by performing experiments with femtosecond pulses in various media (glass, water, and nitrogen gas).  For glass we perform experiments under conditions of normal and anomalous group-velocity-dispersion (GVD) and show that the loss-of-polarization effect is more pronounced in the anomalous-GVD regime where filaments tend to be significantly longer. 

To theoretically explain the loss of polarization in elliptically-polarized, collapsing beams, we consider the nonlinear Schrödinger equations (NLSE) for propagation in a bulk saturable Kerr medium 
\begingroup\makeatletter\def\f@size{9}\check@mathfonts
\begin{equation}
i\frac{\partial A_\pm}{\partial z} = \frac{\partial^2A_\pm}{\partial x^2}+\frac{\partial^2A_\pm}{\partial y^2}+\frac{2}{3}\frac{\left(|A_\pm|^2+2|A_\mp|^2\right)A_\pm}{\left[1+\epsilon\left(|A_\pm|^2+|A_\mp|^2\right)\right]}, \label{eq:nlse}
\end{equation}
\endgroup
where $A_+(z,x,y)$  and $A_-(z,x,y)$ are the slowly-varying envelopes of the clockwise and counter-clockwise circular polarization components of the electric field, $x$ and $y$ are the transverse coordinates normalized by the input beam radius, $z$ is the coordinate along the propagation direction normalized by the diffraction length, and $\epsilon$ is the saturation parameter. The angle $\theta$ between the major axis of the polarization ellipse and the $x$-axis is \cite{sheinfuxMeasuringStabilityPolarization2012}
\begingroup\makeatletter\def\f@size{10}\check@mathfonts
\begin{equation}
\theta (z)=-\frac{1}{2}\tan^{-1}\frac{U}{Q},
\label{eq:eq2}
\end{equation}
\endgroup
where 
\begingroup\makeatletter\def\f@size{9}\check@mathfonts
 $U=-2$ Im$[\int A_+^*A_-rdr]$ and $Q=2$ Re$[\int A_+^*A_-rdr].$\endgroup
 
 An elliptically-polarized Gaussian input beam, whose power $P_{in}$ is moderately above $P_{cr}$, evolves into the coupled spatial solitary waves  
 \begingroup\makeatletter\def\f@size{10}\check@mathfonts
 \begin{equation}
 A_\pm(z,x,y)=e^{i\kappa_\pm Z}R_\pm(x,y), \label{eq:solitary}
 \end{equation}
\endgroup
where $R_\pm$ are solutions of
\begingroup\makeatletter\def\f@size{9}\check@mathfonts
$$-\kappa_\pm R_\pm + \frac{\partial^2R_\pm}{\partial x^2}+\frac{\partial^2R_\pm}{\partial y^2}+\frac{2}{3}\frac{\left(|R_\pm|^2+2|R_\mp|^2\right)R_\pm}{\left[1+\epsilon(|R_\pm|^2+|R_\mp|^2)\right]}=0.$$
\endgroup
When the Gaussian input beam is elliptically-polarized, the power of $A_+(0,x,y)$  is different from that of $A_-(0,x,y)$. Hence, $A_+$ and $A_-$ converge to different solitary waves with $\Delta \kappa = \kappa_+-\kappa_-\neq0,$ and the beam accumulates a polarization angle $\theta_0$ during the initial collapse stage. The polarization rotation angle then satisfies \cite{[{See Supplementary Material}][{ for additional details about simulation and experimental results, experimental setup, calculations, plots and explanations.}]supplementaryLOP2018}
\begingroup\makeatletter\def\f@size{10}\check@mathfonts
\begin{equation}
	\theta(z) = \theta_0 + \frac{\Delta\kappa}{2}z. \label{eq:thetaz}
\end{equation}
\endgroup

In the presence of input noise, $\theta_0$ and $\Delta\kappa$ become random variables, therefore, by the loss of phase lemma \cite{sagivLossPhaseUniversality2017}, the probability distribution of $\theta$ mod $(2\pi)$ converges to a uniform distribution on $[0,2\pi]$ as $z\rightarrow\infty$.  This effect represents a complete \emph{“loss of polarization”}. For $z$ sufficiently large so that $z\Delta P\,(d\Delta\kappa/dz)\gg1$, even for small changes in the input power $\Delta P$, large changes in $\theta$ are induced, making it impossible to deterministically predict the output polarization angle. Note that for a linearly-polarized input beam, since $|A_+|=|A_-|$, both components collapse into identical solitary waves with $\Delta\kappa=0$, and so the polarization angle does not rotate at all. 

\begin{figure}[t]
	\centering
	\includegraphics[width=0.5\textwidth]{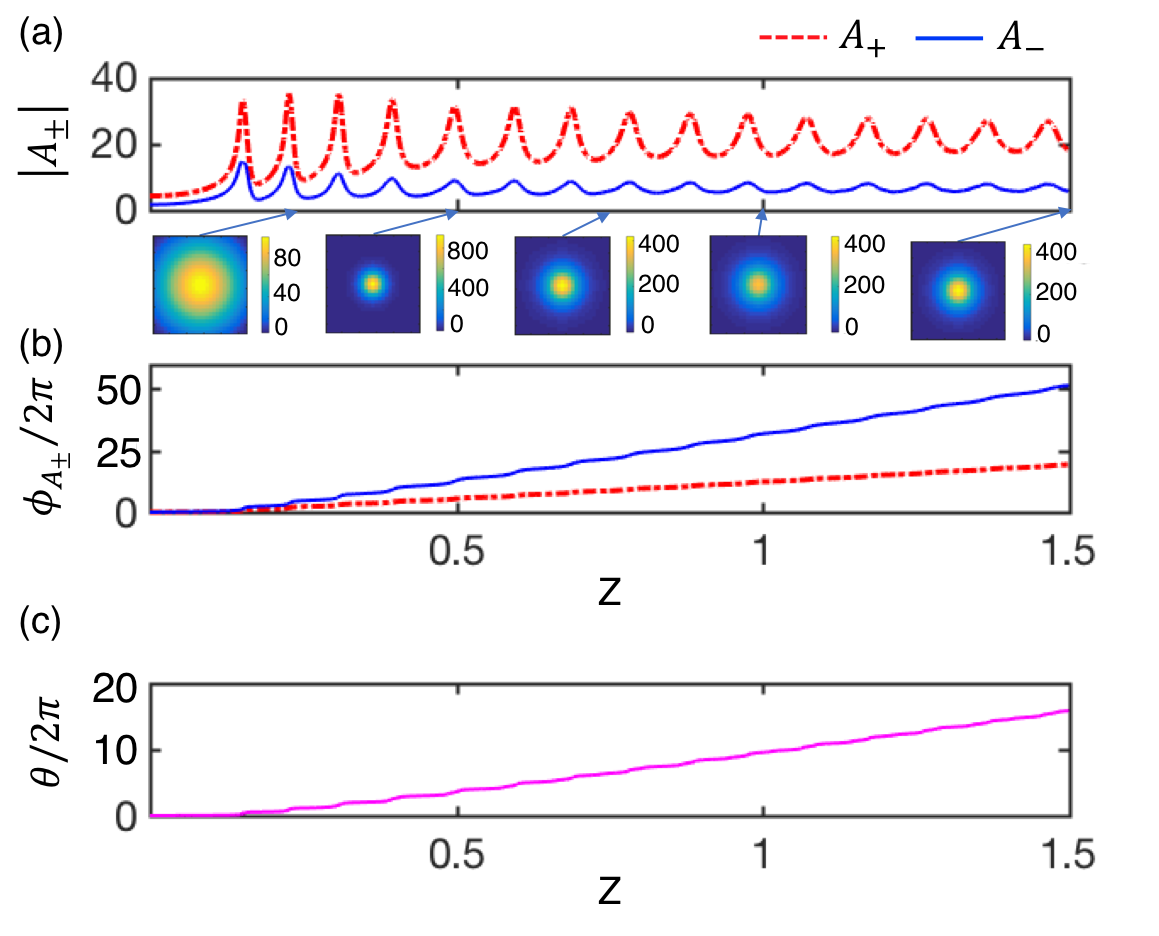}
	\caption{Solution of Eq. (\ref{eq:nlse}) assuming a Gaussian initial beam profile with $P_{in}$/$P_{cr} = 1.73$. (a) On-axis amplitude $|A(z,0)|$ and beam profile vs. propagation distance $z$, (b) Unwrapped on-axis phase $\phi$ for right (left) circularly-polarized component indicated by the red-dotted (blue-solid) curve, (c) The polarization angle $\theta(z)$ vs. $z$.}
	\label{fig:fig1}
\end{figure}

To demonstrate the loss of polarization phenomenon numerically, we solve the coupled NLSE (Eq. (\ref{eq:nlse})) using the split-step Fourier transform method \cite{agrawalNonlinearFiberOptics2013} with $\epsilon = 5\times10^{-5}$ and $|A_+$/$A_-|_{z=0}=2.747$. Both components collapse and evolve into solitary waves, see Fig.~\ref{fig:fig1}(a). The bottom plots of Fig.~\ref{fig:fig1}(a) show the spatial intensity profile of the beam at various propagation lengths. The difference in amplitude between the two components in Fig.~\ref{fig:fig1}(a) corresponds to a difference in their propagation constants, see Eq. (\ref{eq:solitary}). This is shown by the different slopes of the on-axis accumulated phases in Fig.~\ref{fig:fig1}(b) $(\kappa_-\approx 75.82$ and $\kappa_+\approx 26.49)$. Since $\Delta\kappa\approx 49.33$, theoretical prediction of the polarization angle $\theta(z)\approx z\Delta\kappa(P)$/$2\approx 24.66 z$ by Eq. (\ref{eq:thetaz}) agrees well with the direct fit of $\theta(z)\approx 24.47 z$ (Eq. (\ref{eq:eq2})), whose slope is within 0.8\% of the theoretical prediction. To the best of our knowledge, Fig.~\ref{fig:fig1} presents the first example of a multi-component solitary wave of the NLSE with different propagation constants for each component.

\begin{figure}[t]
	\includegraphics[width=0.48\textwidth]{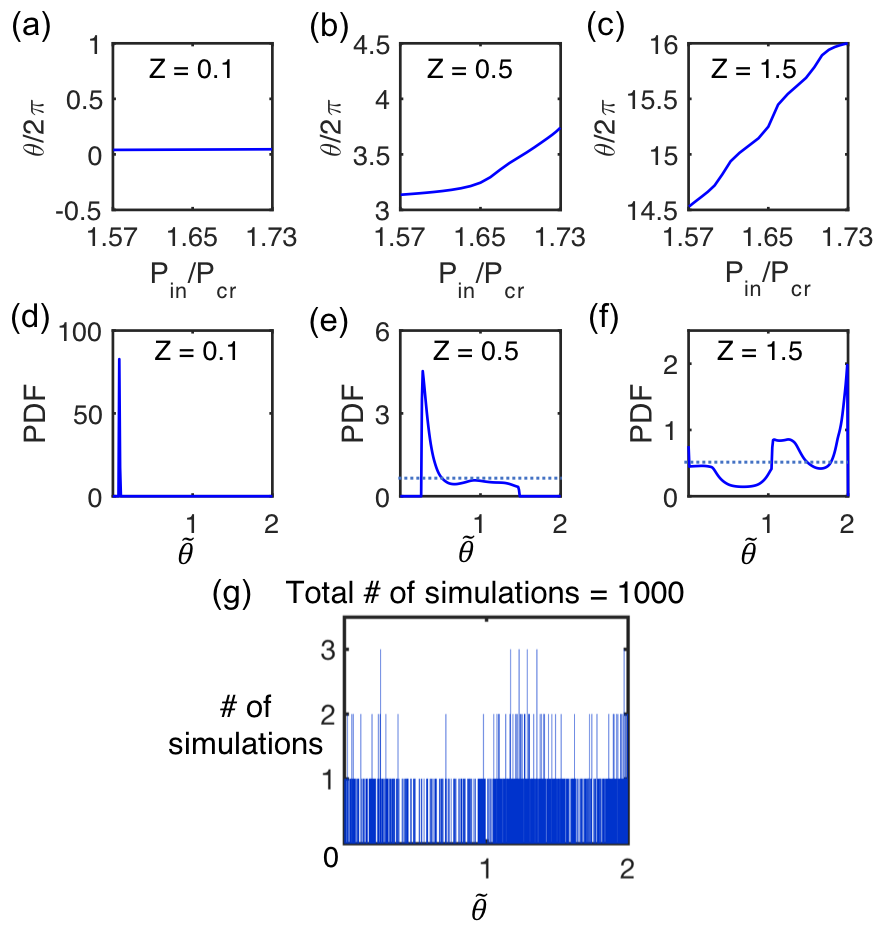}
	\caption{\label{fig:fig2} Elliptically-polarized input beam, (a)-(c): Polarization angle vs. input power at $z$ = 0.1, 0.5 \& 1.5, respectively; (d)-(f): The probability distribution function (PDF) of $\tilde{\theta} = \theta$/$\pi$(mod 2); at $z$ = 0.1, 0.5 \& 1.5; (g): Histogram of $\tilde{\theta}$ at $z$ = 1.5 for 1000 simulations with $P_{in}$/$P_{cr}$ distributed uniformly in [1.53, 1.75].}
\end{figure}

Figure~\ref{fig:fig2} shows the polarization angle $\theta$ as a function of the input power, for various  propagation distances $z$. The elliptically-polarized beam undergoes negligible change in the accumulated polarization angle for short propagation lengths ($z=0.1$) [Fig.~\ref{fig:fig2}(a)]. So the probability distribution function (PDF) is highly localized [Fig.~\ref{fig:fig2}(d)]. As the propagation distance increases ($z=0.5$), these changes increase [Fig.~\ref{fig:fig2}(b)], and the PDF becomes more “spread-out”, i.e., the uncertainty in $\theta$ increases [Fig.~\ref{fig:fig2}(e)]. Ultimately, at long propagation lengths ($z=1.5$), $\theta$ varies rapidly with the power [Fig.~\ref{fig:fig2}(c)], and the PDF approaches a uniform distribution [Fig.~\ref{fig:fig2}(f)]. The PDFs in Figs.~\ref{fig:fig2}(d)-(f) were computed using a novel numerical method, which is more efficient and informative for a fixed number of NLSE simulations than the Monte-Carlo method \cite{ditkowskiSplineBasedApproachUncertainty2018}. We illustrate the loss of polarization at $z=1.5$ by plotting a histogram of 1000 simulations with an elliptically-polarized input beam with $P_{in}$ distributed uniformly in the 10\% interval around 1.65$P_{cr}$ and observe that $\theta$ fluctuates across the entire range of [0,2$\pi$] [Fig.~\ref{fig:fig2}(g)].

As predicted by theory, linearly-polarized beams do not undergo loss of polarization. Indeed, our simulations show that the polarization angle of a linearly-polarized beam remains unchanged irrespective of $P_{in}$ (plot presented in \cite{supplementaryLOP2018}).
 
\begin{figure*}
	\includegraphics[width=1\textwidth]{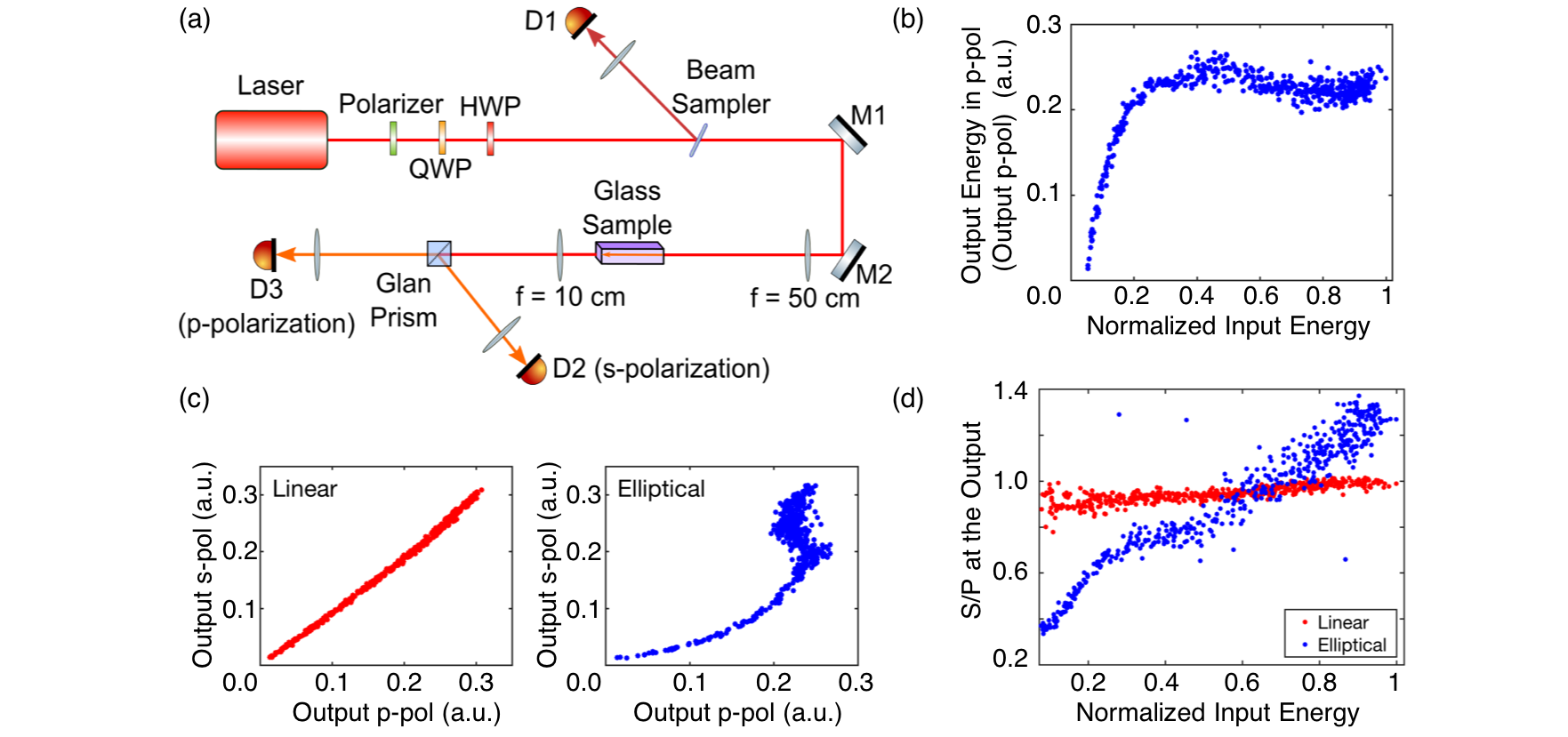}
	\caption{\label{fig:fig3}(a) Experimental setup. QWP: quarter wave plate, HWP: half wave plate, D1, D2, D3: InGaAs detectors (b) Energy in the output $p$-polarization vs. normalized input energy for elliptically-polarized beam. We obtain a similar graph for energy in the output $s$-polarization, (c) Output $s$-polarization vs. $p$-polarization for linearly- (red - left) and elliptically- (blue - right) polarized input, (d) Plot of $s$/$p$ ratio vs. input energy for linearly- (red) and elliptically- (blue) polarized input}
\end{figure*}

 We experimentally investigate the stability of the output polarization after the beam has undergone collapse for different input polarizations (linear and elliptical) in fused silica samples using 1500 nm pulses (75-fs pulse duration, 10-Hz repetition rate)  (further details in \cite{supplementaryLOP2018}). We perform single-shot measurements and record the magnitudes of the $s$- and $p$-polarizations, which can be used to calculate the polarization angle $\theta$ \cite{hechtEllipticalPolarization1998} (see \cite{supplementaryLOP2018} for more details). The experimental setup is shown in Fig.~\ref{fig:fig3}(a). The energy of the input pulses was varied from 15 to 220 $\mu$J.
 
Beam collapse is indicated by the presence of white light at the output due to the generated supercontinuum as a result of filamentation and glass-ionization \cite{gaetaCatastrophicCollapseUltrashort2000,brodeurBandGapDependenceUltrafast1998,brodeurUltrafastWhitelightContinuum1999,yangFilamentationSupercontinuumGeneration2017}. It is also indicated in the plotted curve of output energy in $p$-polarization ($p$-pol) versus input energy [Fig.~\ref{fig:fig3}(b)]. When the input energy is low, the output energy varies linearly with input as expected. When collapse occurs, the transmitted energy saturates due to nonlinear absorption inside the glass sample and the slope of the output energy vs. input energy decreases as shown in Fig.~\ref{fig:fig3}(b). Input energies are normalized to the maximum energy used in our experiments (220 $\mu$J). From both these indicators, we determine that the beam collapse begins at around 0.25 of the normalized energy. 

We plot normalized energies in the $s$- vs. $p$-polarization in Fig.~\ref{fig:fig3}(c) to show trends in $\theta$. For linearly-polarized input, irrespective of whether collapse and filamentation occurs, the curve of $s$- vs. $p$-polarized energy is linear indicating a constant polarization angle. For elliptically-polarized input, in the absence of beam collapse at low powers, the $s$ vs. $p$ plot shows a $\tan^2$ dependence that arises due to steady increase of $\theta$ with power, in agreement with the theory \cite{supplementaryLOP2018}. At high energies, however, $s$ vs. $p$ exhibits random behavior due to loss of polarization. To further investigate this effect, the $s$/$p$ fluence ratio is calculated from each single-shot measurement by taking the ratio of the corresponding detector outputs and is plotted as a function of input energy for elliptically- and linearly-polarized light [Fig.~\ref{fig:fig3}(d)]. The fluctuations in the $s$/$p$ fluence ratio are correlated to the fluctuations in $\theta$. The resulting plot shows two important features.  First, for elliptically-polarized input, the $s$/$p$ ratio steadily increases with input energy, showing a rotation of the polarization ellipse, whereas for linear input the $s$/$p$ ratio stays constant, indicating a constant polarization angle. Second, the fluctuations in $s$/$p$ ratio increase for elliptically-polarized input indicating increased sensitivity of the output polarization angle on input power. For the case of linearly-polarized input, fluctuations in $s$/$p$ ratio remain small and constant throughout. These observations are in accordance with our theoretical prediction based on the nonlinear ellipse rotation phenomenon \cite{boydPropagationIsotropicNonlinear2008}. Below collapse threshold ($< 55\ \mu$J or 0.25 of normalized input energy), the fluctuations for the elliptically-polarized beam are comparable to the small fluctuations ($\sim 1^\circ$) for the linearly-polarized beam. However, for sufficiently high input energy ($> 55\ \mu$J) i.e. when the beam undergoes collapse, the fluctuations become 6 times higher for elliptically-polarized input than for linearly-polarized input, which agrees qualitatively with our numerical predictions. We observe a $27^\circ$ rotation of polarization angle over the entire energy interval in our experiment for elliptically-polarized input. The observed fluctuations in the polarization angle are more than $6^\circ$ for the highest energy in our experiment. This is significantly larger than the measurement uncertainty ($1^\circ$), which we calculate based on the fluctuations in $\theta$ in the linear-polarization case. 

\begin{figure}[t]
	\includegraphics[width=0.48\textwidth]{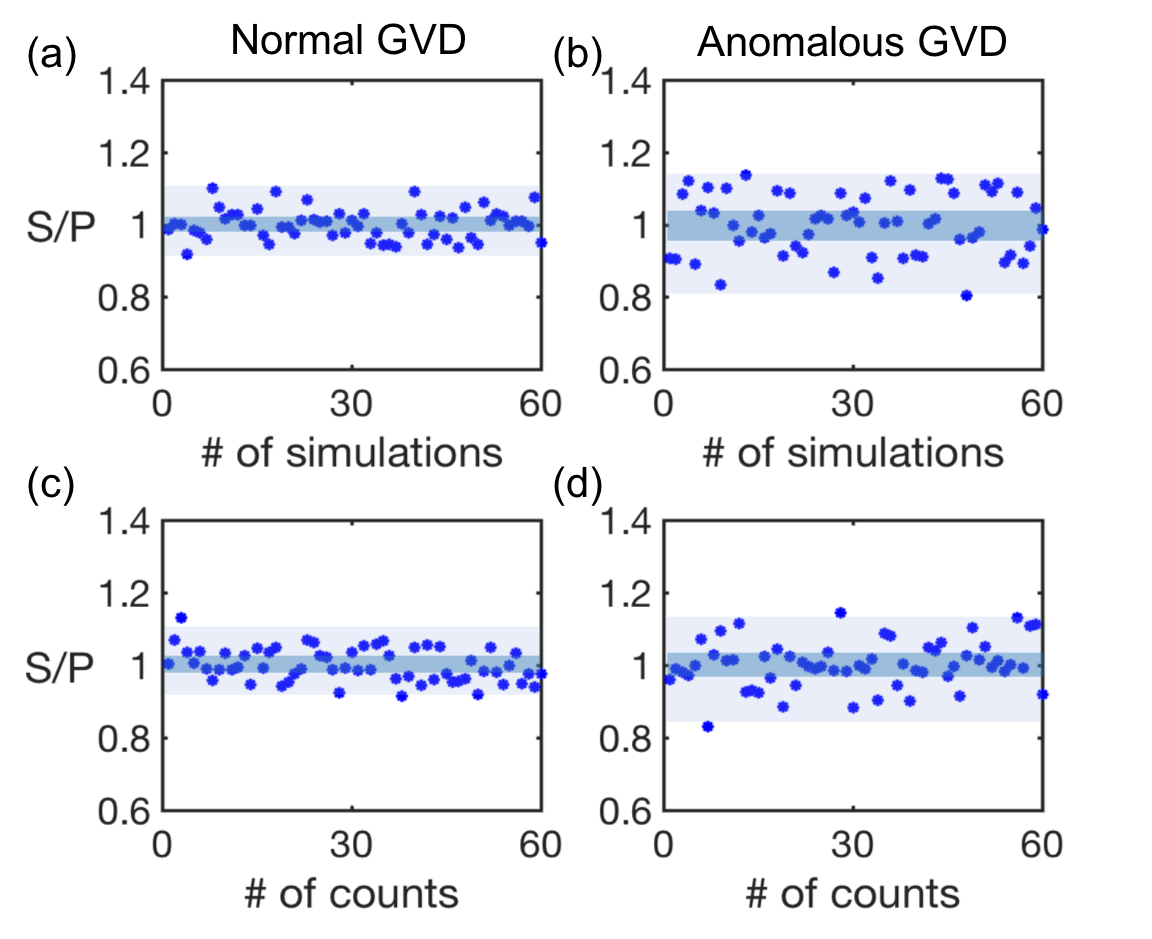}
	\caption{\label{fig:fig4} $s$/$p$ fluence ratios at the output for elliptically-polarized input. Variance ($\sigma$) shown by dark blue shaded region, shot-to-shot fluctuations shown by light blue shaded region. (a),(b): Simulation results with input power varied uniformly between $17.4P_{cr}$  and $19.2P_{cr}$, (c),(d): experimental results in glass at 800 nm and 1500 nm respectively, input energy varied 10\% around (c) 88 $\mu$J and (d) 176 $\mu$J. Experimental results follow the simulation trend that fluctuations are more pronounced in the anomalous-GVD [$\sigma$ = 0.084 (simulation), 0.063 (experiment)] than those in the normal-GVD regime [$\sigma$ = 0.041(simulation), 0.045(experiment)].}
\end{figure}

From our analysis, see Eq. (\ref{eq:thetaz}), and the NLSE simulations [Fig.~\ref{fig:fig1}], we predict as the propagation distance increases, the sensitivity of the output polarization angle to the input power fluctuations increases. Researchers have previously shown that filaments in the anomalous group-velocity dispersion (GVD) regime are longer, more stable and yield more collapsing events as compared to those in the normal-GVD regime \cite{mollRoleDispersionMultiplecollapse2004,durandSelfGuidedPropagationUltrashort2013}. Thus, we expect that the loss of polarization effect would be more prominent in the anomalous-GVD regime. To test this hypothesis, we performed simulations including effects of dispersion, diffraction and nonlinearity for a material with GVD ($\beta_2= \pm26$ ps$^2$/km) similar to glass, 75-fs pulse duration and input power uniformly distributed between $17.4P_{cr}$ and $19.2P_{cr}$. The $s$/$p$ ratio was calculated using the output polarization  angle at $z$ = 0.05. Our simulation results are shown in Figure~\ref{fig:fig4}(a) and (b). For consistency with simulations in the anomalous-GVD regime ($\beta_2= -26$ ps$^2$/km), we use ($\beta_2= +26$ ps$^2$/km) in the normal-GVD-regime simulations. However, our normal-GVD regime experiments were performed with a laser at 800 nm, where $\beta_2$ for glass is slightly different (+35 ps$^2$/km). In the normal-GVD regime, calculated shot-to-shot fluctuations (indicated by light blue shaded region in the plots) in the output polarization angle were $4.9^\circ$; whereas those in the anomalous-GVD regime were about $7.8^\circ$ (1.6 times larger). We also performed corresponding experiments in glass with pulses at 800 nm (normal-GVD regime, $\beta_2= +35$ ps$^2$/km) and at 1500 nm (anomalous-GVD regime, $\beta_2= -26$ ps$^2$/km). Figures~\ref{fig:fig4}(c) and (d) show our experimental results for the normal and anomalous-GVD regime, respectively. Measured shot-to-shot fluctuations in the output polarization angle for the normal-GVD regime were $4.3^\circ$, whereas those in the anomalous-GVD regime were $6^\circ$ (1.4 times larger). Experimental results follow the trend predicted in simulations and confirm our hypothesis that the output polarization angle is more sensitive to the input power in the anomalous-GVD regime than in the normal-GVD regime. In all the plots in Fig.~\ref{fig:fig4}, the input power (energy) is varied by 10\% [$\pm$5\%]. 

This increase in fluctuations occurs in all media whenever there is filamentation of elliptically-polarized beams. To demonstrate this, we performed experiments in glass, liquid water, and in nitrogen gas at 23 bar pressure, using 800 nm, 50-fs pulses at a 10-Hz repetition rate (normal-GVD regime). In all these cases, we compare output $s$/$p$ ratio fluctuations for elliptically-polarized input and linearly-polarized input for fixed fluctuations in input energy. We observed that at low pulse energies (below the collapse threshold), the fluctuations in $s$/$p$ ratio for both input polarizations are identical. On the other hand, above the collapse threshold, fluctuations in $s$/$p$ ratio in case of elliptically-polarized input are 2-4 times higher than the fluctuations in case of linearly-polarized input. (See plots in \cite{supplementaryLOP2018}).

In conclusion, we theoretically show that the loss of polarization angle increases with propagation distance and ultimately leads to a complete loss of polarization angle for collapsing beams of elliptical-polarization. We provide experimental evidence for this effect by measuring a significant increase in the fluctuations of the polarization angle in a glass sample. We demonstrate that the loss of polarization effect is more prominent in the anomalous-GVD regime. Such behavior is universal and should occur in all systems that exhibit multiple collapse of elliptically-polarized beams. Furthermore, this work can be extended to study beam polarization for multi-filamentation. In this case, the loss of polarization effect could lead to spatial de-polarization of the beam due to unequal polarization rotation in each filament. Recent work shows that light with different spatial profiles such as vortex Airy beams and axially asymmetric beams have controllable and designable collapse dynamics that are robust against random noise \cite{chenDynamicControlCollapse2013,chenCollapseDynamicsVector2015,liTamingCollapseOptical2012,liUnveilingStabilityMultiple2016}, and it is expected that the loss-of-polarization effect could also occur in such beams. Our work has implications for applications that depend upon the beam polarization being deterministic for collapsing beams traveling over long distances, such as in filamentation for remote sensing and HHG.

The research of GP, XG, JG and AG was supported by the AFOSR Multidisciplinary University Research Initiative under Award Number FA9550-16-1-0121. The research of AS and GF was partially supported by grant \myhash177/13 from the Israel Science Foundation (ISF).

\end{document}